\documentclass[12pt]{iopart}
\usepackage[pdftex]{color,graphicx}
\begin{document}

\title[Extended hybrid model]{An extended hybrid magnetohydrodynamics gyrokinetic model for numerical simulation of shear Alfv\'en waves in burning plasmas}

\author{X Wang$^{1}$, S Briguglio$^2$, L Chen$^{1,3}$, C Di Troia$^2$, G Fogaccia$^2$, G Vlad$^2$ and F Zonca$^{2,1}$}

\address{$^1$ Institute for Fusion Theory and Simulation, Zhejiang University, Hangzhou, China 310027}
\address{$^3$ Associazione EURATOM-ENEA sulla Fusione, CP 65-00044 Frascati, Roma, Italy}
\address{$^2$ Department of Physics and Astronomy, University of California, Irvine, CA 92697, USA}

\ead{wangxinnku@zju.edu.cn}

\begin{abstract}
Adopting the theoretical framework for the generalized fishbonelike dispersion relation, an extended hybrid magnetohydrodynamics gyrokinetic simulation model has been derived analytically by taking into account both thermal ion compressibility and diamagnetic effects in addition to energetic particle kinetic behaviors. The extended model has been used for implementing an eXtended version of Hybrid Magnetohydrodynamics Gyrokinetic Code (XHMGC) to study thermal ion kinetic effects on Alfv\'enic modes driven by energetic particles, such as kinetic beta induced Alfv\'en eigenmodes in tokamak fusion plasmas.
\end{abstract}

\maketitle
\section{Introduction and motivation}
Nonlinear numerical simulations of magnetohydrodynamics (MHD) and Alfv\'en modes driven by energetic particles (EPs) mostly rely on hybrid MHD gyrokinetic codes, such as HMGC~\cite{briguglio95}, M3D~\cite{park92}, and MEGA~\cite{todo98}. In the hybrid MHD gyrokinetic model, the thermal plasma component is described by MHD, while EP dynamics, in the so-called pressure coupling equation~\cite{park92}, is accounted for via the divergence of the EP pressure tensor, which is computed by solving the gyrokinetic equation with particle in cell (PIC) techniques. Kinetic treatments of the thermal plasma component are well known and generally implemented in (linear) spectral codes, such as NOVA-K~\cite{fu93,gorelenkov99} and MARS-K~\cite{liu08}. More recently, significant developments in gyrokinetic simulation codes, such as GTC~\cite{holod09} and GYRO~\cite{bass10}, have also allowed investigating the kinetic effects of thermal plasma and EP dynamics on long wavelength electro-magnetic fluctuations, which were previously investigated only with codes based on the hybrid MHD-gyrokinetic approach~\cite{briguglio95,park92,todo98,todo05}. In HMGC~\cite{briguglio95}, the thermal plasma description is originally limited to the reduced MHD model~\cite{izzo83}. In the present work, our goal is to extend the hybrid MHD-gyrokinetic model implemented in HMGC~\cite{briguglio95} to the low-frequency domain of the beta induced Alfv\'en eigenmode (BAE) - shear Alfv\'en wave (SAW) continuous spectrum~\cite{chen07a}, where the mode frequency can be generally comparable with thermal ion diamagnetic and/or transit frequencies, i.e. $|\omega|\approx\omega_{\ast pi}\approx\omega_{ti}$. In this frequency range, where kinetic thermal ion (KTI) gap generally exists and influences plasma dynamics~\cite{chen07a}, there is a continuous transition between various MHD and SAW fluctuation branches, as predicted theoretically~\cite{chen84,biglari91,tsai93,chen94,chen95,cheng95,zonca96a,zonca96b,zonca99} and confirmed experimentally~\cite{heidbrink93,heidbrink95,heidbrink99,nabais05,fredrickson06,zonca07,darrow08,podesta09,fredrickson09,zonca09}. Another notable feature of these low frequency fluctuations is that they may be resonantly excited by wave-particle interactions with EPs as well as thermal plasma particles, depending on the perpendicular wavelength~\cite{zonca99,nazikian06}. With the extended hybrid MHD gyrokinetic model discussed here, it will be possible to investigate various problems related with resonant excitation of Alfv\'enic and MHD fluctuations by EPs in the BAE-SAW continuous spectrum, consistent with gyrokinetic codes, e.g., GTC~\cite{lin98}, in a common validity domain. Therefore, both the eXtended HMGC (XHMGC) and GTC codes can be verified using different models; yielding more detailed understanding of the underlying physics. In fact,  theoretical and numerical work, presented in this article and partly developed within the framework of the SciDAC project on ``Gyrokinetic Simulation of Energetic Particle Turbulence and Transport'' (GSEP), was the prerequisite for successful verification of XHMGC predictions against analytic theories~\cite{wang10a} as well as GTC numerical simulation results~\cite{deng10,zhang10}, reported recently.

In this work, we extend the hybrid MHD-gyrokinetic model, derived originally in~\cite{park92} for applications to numerical simulations of EP driven Alfv\'en modes. The main differences with respect to the usual pressure coupling equation~\cite{park92} are due to renormalization of the inertia term, to properly account for finite thermal ion diamagnetic effects, as well as to the gyrokinetic treatment of the thermal ion pressure tensor, which allows us to properly handle wave-particle resonant interactions in the low frequency regime, where they can be of crucial importance for the analysis of linear and nonlinear behaviors of collisionless burning plasmas. The extended model has been developed assuming ideal Ohm's law as well as ignoring finite Larmor radius (FLR) effects in order to simplify the technical complications while still maintaining all essential physics ingredients~\cite{zonca06}. In practice, maintaining the ideal MHD Ohm's law as limiting case implies assuming $T_e\ll T_i$ and neglecting ion FLR effects, although finite magnetic drift orbit widths (FOW) are fully retained~\cite{wang10b}. A more general approach without these simplifying assumptions will be developed in a separate work. For demonstrating the validity of the modified equations, we show that they are equivalent to the quasi-neutrality and vorticity equations derived in~\cite{zonca06} for the frequency range from the kinetic ballooning mode (KBM) and BAE to the toroidal Alfv\'en eigenmode (TAE). The XHMGC model equations in the linear limit are equivalent to the extended kinetic MHD used in spectral codes, such as NOVA-K~\cite{fu93,gorelenkov99} and MARS-K~\cite{liu08}, but with EP dynamics treated non-perturbatively and on the same footing as the thermal plasma response (see Section~\ref{eqs} for more details). The possibility of investigating nonlinear dynamics, however, makes XHMGC more suitable to direct comparisons with M3D~\cite{park92} or gyrokinetic codes~\cite{lin98,waltz02} in a common validity domain.

The paper is organized as follows. In Section~\ref{eqs}, the extended hybrid model equations are presented and discussed within the theoretical framework of~\cite{zonca06}. In Section~\ref{nuimp},  we describe the numerical implementation of the extended model into HMGC, by adding both thermal ion compressibility and diamagnetic effects (of thermal ions as well as EPs) into MHD equations and a thermal ion population in the PIC module. In Section~\ref{app}, possible applications and validity limits of XHMGC are discussed. A synthetic summary of current BAE numerical simulation results~\cite{wang10a} are also provided. Finally, conclusions and discussions are given in Section~\ref{conclusions}. 
\section{Derivation of the extended hybrid model}\label{eqs}
Reference~\cite{zonca06} presents a general theoretical framework for stability analyses of various modes and the respective governing equations. It shows that all modes of the shear Alfv\'en branch having frequencies in the range between the thermal ion transit and Alfv\'en frequency can be consistently described by one single general fishbone-like dispersion relation (GFLDR)~\cite{chen84,chen95,zonca06,chen94, tsai93}. Reference~\cite{zonca06} discusses various reduced equations governing the evolution of SAW fluctuations in burning plasmas, using the general approach of reference~\cite{chen91}. In this sense, reference~\cite{zonca06} could seem not the optimal eliminate reference framework for further generalizing the HMGC hybrid model equations ~\cite{briguglio95,briguglio98}, which are to be used for nonlinear studies as well. However, the detailed analyses of reduced model equations, reported in~\cite{zonca06}, on the basis of specializations of ordering of dimensionless parameters in the case of burning plasmas of fusion interest, starting from the somewhat different orderings of interest to space plasmas given in~\cite{chen91}, allow to fully grasp the physics implications of the underlying approximations. Moreover, on the basis of our discussions, it is straightforward to motivate the extension of the derived model equations to the nonlinear case, as shown at the end of this section.

Considering that the characteristic frequency, $|\omega|$, is much lower than the ion cyclotron frequency, $|\omega_{ci}|$, we may adopt the gyrokinetic theoretical approach and closely follow reference~\cite{chen91}. The low-frequency plasma oscillations can, thus, be described in terms of three fluctuating scalar fields: the scalar potential perturbation $\delta\phi$, the parallel (to ${\bf b} = {\bf B_0} /B_0$, with ${\bf B_0}$ the equilibrium magnetic field) magnetic field perturbation $\delta B_{\parallel}$ and the perturbed field $\delta\psi$, which is related to the parallel vector potential fluctuation $\delta A_{\parallel}$ by
\begin{equation}\label{eq1}
\delta A_{\parallel}\equiv -i\left(\frac{c}{\omega}\right){\bf b}\cdot\nabla\delta\psi.
\end{equation}
The governing equations for describing the excitation of the shear Alfv\'en frequency spectrum by energetic ions precession, precession-bounce and transit resonances in the range $\omega_{\ast pi}\approx\omega_{ti}\leq\omega\leq\omega_A$, covering the entire frequency range from KBM/BAE~\cite{tsai93,biglari91,heidbrink93,turnbull93} to TAE~\cite{cheng85,chen88,fu89}, are generalized kinetic vorticity equation and quasi-neutrality condition, which can be written as followings, in the limit of vanishing FLR (see equation~(16) and equation~(17) in reference~\cite{zonca06}):
\begin{eqnarray}\label{0616}
{\bf B_0}\cdot\nabla\left(\frac{k_{\perp}^{2}}{k_{\theta}^{2}B_0^{2}}{\bf
B_0}\cdot\nabla\delta\psi\right)
+\frac{\omega(\omega-\omega_{*pi}- \frac{n_E m_E}{n_i m_i}\omega_{*pE})}{v_{A}^{2}}\frac{k_{\perp}^{2}}{k_{\theta}^{2}}\delta\phi\nonumber\\
-\left<\sum_{s\neq e}\frac{4\pi
e_{s}}{k_{\theta}^{2}c^{2}}\omega\hat{\omega}_{ds}\delta
K_{s}\right>
+\sum_{s}\frac{4\pi}{k_{\theta}^{2}B_0^{2}}{\bf k}\times{\bf
b}\cdot\nabla
(P_{s\perp}+P_{s\parallel})\Omega_{\kappa}\delta\psi=0,
\end{eqnarray}
\begin{equation}\label{0617}
\left\langle\sum_{s\ne E}\frac{e^2_s}{m_s}\frac{\partial F_{0s}}{\partial\varepsilon}\right\rangle(\delta\phi-\delta\psi)+\sum_{s=i}e_s\langle\delta K_s\rangle=0,
\end{equation}
where the non-adiabatic particle response, $\delta K_s$, is obtained via the drift-kinetic equation
\begin{eqnarray}\label{gyroeq}
\lbrack\omega_{tr}\partial_{\theta}-i(\omega-\omega_d)\rbrack_s\delta K_s = i\left(\frac{e}{m}\right)_sQF_{0s}\left\lbrack(\delta\phi-\delta\psi)
+\left(\frac{\hat{\omega}_d}{\omega}\right)_s\delta\psi\right\rbrack.
\end{eqnarray}
Here, angular brackets stand for velocity space integration, $s$ denotes all particle species ($e=$ bulk electrons, $i=$ bulk ions, $E=$ energetic particles), $e_s$ and $m_s$ are the species electric charge and mass, $F_{0s}$ is the equilibrium distribution function (generally anisotropic), $\varepsilon=v^2/2$ the energy per unit mass, $QF_{0s}=(\omega\partial_{\varepsilon}+\hat\omega_{\ast})_{s}F_{0s}$, $\hat{\omega}_{*s}F_{0s}=\omega_{cs}^{-1}{\bf(k\times b)}\cdot\nabla F_{0s}$, ${\bf k}\equiv-i\nabla$ is the wave vector, $\omega_{cs}=e_sB_0/m_sc$ is the cyclotron frequency, $k_{\perp}$ is the perpendicular wave vector, $\omega_{\ast ps}=({\bf k}\times{\bf b}\cdot\nabla P_s)/n_sm_s\omega_{cs}$ is the diamagnetic frequency, $P_{s\perp}$ and $P_{s\parallel}$ are, respectively, the total perpendicular and parallel plasma pressures, $\omega_{tr}=v_{\parallel}/qR$ is the transit frequency and $\hat{\omega}_{ds}=(m_sc/e_s)(\mu+v^2_{\parallel}/B_0)\Omega_\kappa$, with $\Omega_\kappa=\bf{k}\times{\bf b}\cdot{\bf\kappa}$ and ${\bf\kappa}={\bf b}\cdot\nabla{\bf b}$. Note that the difference between $\hat{\omega}_{ds}$ and $\omega_{ds}=(m_sc/e_s)(\mu\Omega_B+v^2_{\parallel}\Omega_\kappa/B_0)$, with $\Omega_B={\bf k}\times{\bf b}\cdot\nabla B_0/B_0$, has been discussed in~\cite{zonca06,chen91} and, generally, must be handled properly; although, for many applications in low pressure ($\beta=8\pi P/B_0^2\ll1$) plasmas, one can consider $\omega_{ds}=\hat{\omega}_{ds}$ after solving for $\delta B_{\parallel}$ from perpendicular pressure balance~\cite{chen91,zonca06}, as implicitly assumed in equations~\ref{0616},~\ref{0617} and~\ref{gyroeq}. Note, also, that we have maintained the EP contribution to the divergence of the polarization current, which is represented by its leading term $\propto\omega_{*pE}$ in equation~\ref{0616}. This term is readily derived from the last term on the left hand side (LHS) of equation (13) in reference~\cite{zonca06} (see also \ref{AA} for further details) and was neglected in there due to the ordering $\beta_E /\beta_b\approx\tau_E/\tau_{SD} < 1$, valid in a burning plasma dominated by fusion alpha particle self-heating. Here, $\beta_E$ and $\beta_b$ denote the beta values of EP and bulk plasma components (electrons and thermal ions), respectively, while $\tau_E$ and $\tau_{SD}$ are the energy confinement time and EP slowing down time. More generally~\cite{chen84,biglari91,tsai93,chen94}, the ordering $\beta_E\approx\beta_b$ better represents nowadays magnetized plasmas of fusion interest and, thus, $n_E \omega_{*pE} \approx n_i \omega_{*pi}$, as assumed in equation~\ref{0616}.

Equations~\ref{0616} and~\ref{0617}, together with the drift-kinetic equation, equation~\ref{gyroeq}, are the simplest yet relevant equations for analyzing the resonant excitations of SAW by EPs. Equation~\ref{0616} demonstrates that both resonant as well as non-resonant responses due to the $\propto\delta K_E$ term enter via the magnetic curvature drift coupling. In the high frequency case, $\omega_A\ge\omega\ge\omega_{\ast pi}\gg\omega_{ti}$, the thermal ion kinetic compression response $\delta K_i$ can be neglected. Thus, the quasi-neutrality condition, equation~\ref{0617}, reduces to the ideal MHD approximation, $\delta\phi\simeq\delta\psi$; i.e. $\delta E_{\parallel}\simeq 0$~\cite{chen91}. Meanwhile, neglecting the $\propto\omega_{*pi}, \omega_{*pE}$ terms, equation~\ref{0616} becomes equivalent to equation~(3) in~\cite{park92}, i.e. the following pressure coupling equation in the hybrid MHD-gyrokinetic approach
\begin{equation}\label{PC}
\rho_{b}\frac{d{\bf v_{b}}}{dt}=-\nabla P_b-(\nabla\cdot{\bf P_{E}})_{\perp}+\frac{{\bf J}\times{\bf B}}{c};
\end{equation}
where the subscript $b$ denotes the bulk plasma (electrons and thermal ions), while $\rho_b$ and ${\bf v}_b$ are, respectively, bulk plasma mass density and fluid velocity. Here, the EP contribution to the perpendicular momentum change of the plasma has been neglected, due to $n_E/n_b\ll|\omega/\omega_{\ast E}|$~\cite{zonca06,park92}, and thermal ion diamagnetic effects are consistently dropped since $n_E \omega_{*pE} \approx n_i \omega_{*pi}$.

In order to extend the hybrid model to the low-frequency regime where $\omega\sim\omega_{ti}$, we need to include the effects of the thermal ion compressibility within the hybrid simulation scheme. That is, we need to include effects associated with the $\delta K_i$ terms in equations~\ref{0616} and \ref{0617}. First, in order to simplify the discussions, we formally assume $T_e/T_i\rightarrow 0$ in the present work; the general case with finite $T_e$ will be considered elsewhere. Thus, according to equation~\ref{0617}, we have $\delta\phi-\delta\psi\simeq0$ and the ideal MHD condition $\delta E_{\parallel}\simeq0$ remains valid. Next, we proceed to establish correspondences between the pressure coupling equation, equation~\ref{PC}, and the generalized kinetic vorticity equation, equation~\ref{0616}.

Applying the operator $(\partial/\partial t){\bf\nabla}\cdot({\bf B_0}/B_0^2)\times$ to the linearized equation~\ref{PC} and noting the quasi-neutrality condition $\nabla\cdot{\bf J}=0$, we readily derive
\begin{eqnarray}\label{1}
\underbrace{\frac{1}{c}\frac{\partial}{\partial t}{{\bf
B}_{0}}\cdot\nabla\frac{\delta
J_{\parallel}}{B_{0}}}_{i}+\underbrace{\frac{1}{c}\frac{\partial}{\partial t}\delta{\bf B}\cdot\nabla\left(\frac{J_{\parallel0}}{B_0}\right)}_{ii}+\underbrace{\frac{\partial}{\partial
t}\nabla\cdot\left(\frac{{{\bf B}_{0}}\times\rho_{b0}\frac{d{{\bf\delta
v}_{b}}}{dt}}{B_{0}^{2}}\right)}_{iii}\nonumber\\
+\underbrace{\frac{\partial}{\partial
t}\nabla\delta P_{b}\cdot\left(\nabla\times\frac{{{\bf b}}}{B_{0}}\right)}_{iv}
+\underbrace{\frac{\partial}{\partial
t}\nabla\cdot\left(\frac{{\bf b}\times(\nabla\cdot{\bf\delta
P_{E}})_{\perp}}{B_{0}}\right)}_{v}=0.
\end{eqnarray}
Noting also the parallel Amp\`ere's law along with ${\bf\nabla}\cdot\delta{\bf A}=0$,
\begin{equation}
4\pi\delta J_{\parallel}=-c\nabla^2\delta A_{\parallel},
\end{equation}
and equation~\ref{eq1}, term (i) can be seen to correspond to the field line bending term; i.e. the first term in equation~\ref{0616}. Term (ii), on the contrary, does not have any direct correspondence in equation~\ref{0616}. This term is the usual kink drive and it was dropped in the analysis of~\cite{zonca06}, focusing on drift Alfv\'en fluctuations with high mode numbers, for it is formally of $O(1/n)$, with $n$ the toroidal mode number. However, as noted in equation~(A1) of~\cite{zonca06}, term (ii) is readily recovered in a form that can be straightforwardly reduced to that reported here. Meanwhile, from the linearized Ohm's law
\begin{equation}
{\bf\delta E_{\perp}}+\frac{1}{c}{\bf\delta v_b}\times{\bf B_0}=0,
\end{equation}
and ${\bf\delta E_{\perp}}=-\nabla_{\perp}\delta\phi$, term (iii) corresponds to the second term in equation~\ref{0616} with the $\propto\omega_{*pi}, \omega_{*pE}$ terms neglected. To establish correspondences between the pressure responses in equations~\ref{0616} and \ref{1}, we first denote $P_b=P_e+P_i$. It can then be shown (see appendix \ref{A}) that term (iv) corresponds to the thermal ion and electron contributions to the last term on the LHS of equation~\ref{0616}, when kinetic compression effects of the background thermal plasma are neglected.

Finally, let us discuss term (v), due to EP pressure perturbation, which can be expressed as (see \ref{B}). 
\begin{eqnarray}\label{B5}
\frac{\partial}{\partial t}\nabla\cdot\left(\frac{{\bf
b}\times(\nabla\cdot{\bf\delta
P_{E}})_{\perp}}{B_{0}}\right) =\frac{\omega}{B_{0}}\Omega_{\kappa}(\delta P_{E\parallel}+\delta
P_{E\perp}).
\end{eqnarray}
Meanwhile, noting the definition of $\delta K_s$~\cite{chen91, zonca06}, the $\delta K_s$ term in equation~\ref{0616} can be shown to be related with the pressure perturbations as (see \ref{C})
\begin{eqnarray}\label{deltakE}
\left\langle\frac{4\pi
e_{s}}{k^{2}_{\theta}c^{2}}\omega\hat{\omega}_{ds}\delta
K_{s}\right\rangle&=&\frac{4\pi\omega}{k^{2}_{\theta}cB_0}\Omega_{\kappa}(\delta
P_{s\perp}+\delta P_{s\parallel})\nonumber\\
&&-\left \langle \frac{4\pi e_s^2}{k_\theta^2 m_s c^2} \omega \hat\omega_{ds} \frac{\partial F_{0s}}{\partial \varepsilon}\right\rangle \left( \delta \phi - \delta \psi \right)\nonumber\\
&&+\frac{4\pi}{k_{\theta}^{2}B_0^{2}}({{\bf k}\times{\bf b}})\cdot(\nabla P_{0s\perp}+\nabla
P_{0s\parallel})\Omega_{\kappa}\delta\psi.
\end{eqnarray}
Note that the 2nd term in the right hand side (RHS) disappears in the ideal MHD $\delta\phi\simeq\delta\psi$ limit. Equation~\ref{deltakE}, thus, clearly demonstrate that the $\delta K_s$ contribution in equation~\ref{0616}, combined with the 3rd term on the RHS of equation~\ref{deltakE} (or the last term on the LHS in equation~\ref{0616}), has the same form of equation~\ref{B5} and recovers the total pressure response of term (iv) in equation~\ref{1} for $\delta P_{\perp i}=\delta P_{\parallel i}=\delta P_i$. In other words, the $\delta K_s$ term in equation~\ref{0616} corresponds to the kinetic compressibility component of the pressure perturbations.

Summarizing the above discussions, it is clear that, in order to include effects due to finite thermal ion compressibility and diamagnetic drift as well as the finite EP contribution to the divergence of the polarization current, the pressure coupling equation in the MHD-gyrokinetic approach, equation~\ref{PC}, has to be modified such that its perpendicular components are given by equation~\ref{eq:twofluidperp} of \ref{AA}, which we rewrite here for the reader's convenience
\begin{eqnarray}\label{eq:twofluidperp2}
& & \left[ \rho_b \left(\frac{\partial}{\partial t} + {\bf v}_b \cdot {\bf \nabla} \right) + \frac{{\bf b} \times {\bf \nabla} P_{0 E \perp}}{\omega_{cE}} \cdot {\bf \nabla} \right]  \delta {\bf v}_b = \nonumber \\ & & \hspace*{3em} - {\bf \nabla}_\perp P_e - \left( {\bf \nabla} \cdot {\bf P}_i \right)_\perp -  \left( {\bf \nabla} \cdot {\bf P}_E\right)_\perp + \left(\frac{{\bf J} \times {\bf B}}{c} \right)_\perp \;\; . 
\end{eqnarray}
Here, ${\bf v}_b = {\bf b} \times {\bf \nabla} P_{0 i \perp}/(\rho_b \omega_{ci}) + \delta {\bf v}_b$, $\delta {\bf v}_b = (c/B_0) \delta {\bf E} \times {\bf b}$ and the ``unshifted'' pressure tensors ${\bf P_E}$ and ${\bf P_i}$ need to be calculated from solutions of the gyrokinetic equations as specified in \ref{AA}, while $P_e$ is consistently neglected in the present approach, assuming $T_e/T_i\rightarrow 0$. Reminding the concluding remark of \ref{AA}, this equation readily reduces to the well-known pressure coupling equation~\ref{0616}, in the limit where thermal ion diamagnetic effects and EP contribution to the divergence of the polarization current are neglected.

As anticipated above, in the present work, we followed reference~\cite{zonca06}, since that has a detailed discussion of validity limits of different reduced models of the whole vorticity and quasi-neutrality equations, derived for fusion applications and following the trace of reference~\cite{chen91}. Equation~\ref{eq:twofluidperp2} includes equilibrium parallel current effects, as discussed earlier in this section and in~\cite{zonca06} (Appendix). This simple remark readily follows from the discussion presented in~\cite{deng10} as well as the modified momentum balance equation implemented in XHMGC, i.e. equation~\ref{eq:twofluidperp2} itself.
The present model is valid in the nonlinear case too, as shown by the simple derivation provided in \ref{AA} and by the following discussion.  This is deduced easily from direct inspection of equation~(5) in reference~\cite{chen01}. That equation clearly shows that, for the small FLR limit considered in HMGC~\cite{briguglio95,briguglio98}, the nonlinear terms, treated explicitly, are those that are coming from convective ${\bf E}\times{\bf B}$ nonlinearity and from the Maxwell stress nonlinearity, when the thermal ion response is taken in the fluid limit, both of which are readily obtained from equation~\ref{eq:twofluidperp2} upon application of the operator $\partial_t\nabla\cdot({\bf B_0}/B_0^2)\times$, as it was done for equation~\ref{PC} earlier in the section. Other nonlinear dynamics, which are implicitly included in $(\nabla\cdot{\bf P_i})$ and $(\nabla\cdot{\bf P_E})_{\perp}$ terms, are fully retained via equation~\ref{eq:twofluidperp2}. Thus, the back reaction of zonal structures onto SAW fluctuations is fully accounted for, i.e. that  of zonal flows (ZFs) and fields as well as radial modulation of equilibrium profiles~\cite{bass10,chen07a,chen07b} which also enter via the diamagnetic terms in equation~\ref{eq:twofluidperp2}, computed on the whole (slowly evolving) thermal ion and EP pressure profile, obtained from the respective toroidally and poloidally averaged distribution functions. This choice is consistent with known approaches to nonlinear MHD equations, accounting for finite diamagnetic drift corrections~\cite{tang80,tang82,sugiyama00,mizuguchi00,hazetline92}. 

Thus, the approximations involved with the extended implementation within XHMGC on the basis of equation~\ref{eq:twofluidperp2} consist of neglecting FLR, assuming electron as a massless fluid, considering $T_e/T_i\rightarrow0$ (such that parallel Ohm's law is recovered in the ideal MHD limit) and accounting for Reynolds stress in the thermal ion fluid limit. The possible further extension of the present model to include finite $T_e/T_i$ and generalizing the parallel Ohm's law, while maintaining other simplifying assumptions, is straightforward on the basis of the present discussion and will be reported in a separate work~\cite{wang10c}. Here we note that the present extended hybrid model, based on equation~\ref{eq:twofluidperp2}, with clearly formulated assumptions that limit its applicability, includes very rich physics; e.g. it is capable to correctly evaluate the renormalized inertia for ZFs, for which the trapped thermal ion dynamics is of crucial importance, and to account for geodesic acoustic mode (GAM) kinetic response, including Landau damping. 

So far, XHMGC has been used for moderate EP drive~\cite{wang10a,deng10,zhang10,wang10b}, where the EP diamagnetic correction to the divergence of the polarization current can be neglected, as argued in~\cite{zonca06}. Actually, in the studies reported in~\cite{wang10a}, thermal ion diamagnetic contribution to the polarization current is also neglected, since the case of uniform thermal ion pressure profiles is investigated in there for facilitating comparisons of numerical simulation  results with analytic theory predictions (see also section~\ref{app}).
\section{Numerical implementation}\label{nuimp}
HMGC~\cite{briguglio95} is used for investigating linear and nonlinear properties of moderate toroidal number (n) shear Alfv\'en modes in tokamaks. It solves the coupled set of O($\epsilon^3$) reduced-MHD equations~\cite{izzo83} for the electromagnetic fields and the gyro-center Vlasov equation for a population of energetic ions, where large aspect ratio is assumed, i.e. $\epsilon=a/R_0\ll1$, with $a$ and $R_0$ the tokamak minor and major radius, respectively . Energetic particles contribute to the dynamic evolution of the wave fields via the pressure tensor term in the MHD equations, as described by the pressure coupling equation~\cite{park92}. This code allows us to describe both self-consistent mode structures in toroidal equilibria and EP dynamics, as well as to get a deeper insight into how the Alfv\'enic modes affect the confinement of such particles.

The extended model, described in section~\ref{eqs}, has been implemented into the eXtended version of HMGC (XHMGC). Following the general procedure, described in references~\cite{briguglio95,briguglio98}, for the formal manipulation of equation~\ref{eq:twofluidperp2}, the relevant equations for the MHD solver are in terms of the poloidal magnetic field stream function $\Psi$ and $U$, which is proportional to the scalar potential $\Phi$ and defined as $U=-c\Phi/B_0$, can be written in the following form in the cylindrical coordinate system $(R, Z, \varphi)$:
\begin{eqnarray}\label{S1}
\frac{\partial\Psi}{\partial t}=\frac{R^2}{R_0}\nabla\Psi\times\nabla\varphi\cdot\nabla U+\frac{B_0}{R_0}\frac{\partial U}{\partial\varphi}+\eta\frac{c^2}{4\pi}\triangle^\ast\Psi+O(\epsilon^4v_AB_\varphi),
\end{eqnarray}
\begin{eqnarray}\label{S2}
& & \hat\rho\left(\frac{D}{Dt}+\frac{2}{R_0}\frac{\partial U}{\partial Z}\right)\nabla^2_{\perp}U+\nabla\hat\rho\cdot\left(\frac{D}{Dt}+\frac{1}{R_0}\frac{\partial U}{\partial Z}\right)\nabla_\perp U\nonumber\\
& &-\left( \frac{R^2}{R_0^3\omega_{ci0}} \frac{\partial P_{0i\perp}}{\partial Z} + \frac{R^2}{R_0^3\omega_{cE0}} \frac{\partial P_{0E\perp}}{\partial Z}\right) \nabla_\perp^2 U \nonumber\\
& &- {\bf \nabla} \left( \frac{R^2}{R_0^3\omega_{ci0}} \frac{\partial P_{0i\perp}}{\partial Z} + \frac{R^2}{R_0^3\omega_{cE0}} \frac{\partial P_{0E\perp}}{\partial Z} \right) \cdot {\bf \nabla}_\perp U\nonumber\\
& &+\nabla\cdot\left\lbrack\frac{R^4}{R_0^3}\left(\nabla\varphi\times\frac{\nabla P_{0E\perp}}{\omega_{cE0}}+\nabla\varphi\times\frac{\nabla P_{0i\perp}}{\omega_{ci0}}\right)\cdot\nabla\nabla_{\perp}U\right\rbrack\nonumber\\
& = &\frac{1}{4\pi}{\bf B}\cdot\nabla\triangle^\ast\Psi+\frac{1}{R_0}\nabla\cdot\lbrack R^2(\nabla P_e+\nabla\cdot\Pi_i+\nabla\cdot\Pi_E)\times\nabla\varphi\rbrack\nonumber\\
& &+O\left(\epsilon^4\rho\frac{v^4_A}{a^2}\right),
\end{eqnarray}
where we have maintained the same notation of reference~\cite{briguglio95} and explicitly show the additional terms that have been added to implement the extended XHMGC model. Thus, 
\begin{eqnarray*}
\hat\rho=\frac{R^2}{R^2_0}\rho,\ \ \frac{D}{Dt}=\frac{\partial}{\partial t}+\frac{R^2}{R_0}\nabla U\times\nabla\varphi\cdot\nabla, \nonumber\\
\nabla_{\perp}^2\equiv\frac{1}{R}\frac{\partial}{\partial R}R\frac{\partial}{\partial R}+\frac{\partial^2}{\partial Z^2},
\end{eqnarray*}
the Grad-Shafranov operator $\triangle^\ast$ is defined by
\begin{equation}
\triangle^\ast\equiv R\frac{\partial}{\partial R}\frac{1}{R}\frac{\partial}{\partial R}+\frac{\partial^2}{\partial Z^2},
\end{equation}
$B_0$ is the vacuum magnetic field on the magnetic axis at $R=R_0$~\footnote{Please, note the difference between the present notation, where $B_0$ stands for the on axis equilibrium magnetic field, and that used in section II, where $B_0$ generally denoted the (spatially dependent) equilibrium magnetic field.}, $\omega_{ci0}=e_i B_0/(m_i c)$, $\omega_{cE0}=e_E B_0/(m_E c)$ and the subscript $\perp$ denotes components perpendicular to $\varphi$. In the above equations, ${\bf v_\perp}$ is the ${\bf E} \times {\bf B}$ fluid velocity, $\rho$ is the bulk plasma mass density, $P_e$ is the scalar pressure of bulk electrons, $\Pi_E$ and $\Pi_i$ are, respectively, the pressure-tensor of the EP and thermal ions, computed with the definitions of equations~\ref{eq:pperp} and \ref{eq:pparallel}, given in \ref{AA}, and, $\eta$ is the resistivity and $c$ is the speed of light. These $O(\epsilon^3)$ equations have been first derived in reference~\cite{izzo83}, limited to the MHD description of the thermal plasma, while the inclusion of energetic particle dynamics has been discussed in references.~\cite{briguglio95,briguglio98}. Here, in the proposed further extension of the numerical model, thermal ion dynamics as well as diamagnetic effects are also taken into account, according to equation~\ref{eq:twofluidperp2}, derived in \ref{AA} and section~\ref{eqs}. At the leading order in $\epsilon$, $O(\epsilon^2)$, the reduced-MHD equations describe the thermal plasma in the cylindrical approximation. Toroidal geometry enters the equations as corrections at the next order in the inverse aspect ratio.

In order to close equations~\ref{S1} and~\ref{S2}, the EP and thermal ion pressure tensor components can be obtained by directly calculating the appropriate velocity moments of the distribution function for the particle population interacting with the perturbed electromagnetic field. As discussed in section~\ref{eqs}, we initially assume the $T_e/T_i\rightarrow0$ limit for the sake of simplicity, i.e. $P_e\rightarrow0$. Meanwhile, with cold electron assumption and ignoring thermal ion finite Larmor radius (FLR), ideal MHD parallel Ohm's law can be readily recovered.

As to numerical formulation, the equations of motion in gyro-center coordinates for thermal ions are in the same form, mutatis mutandis, as those reported in  \cite{briguglio95} for EPs. In the gyrocenter-coordinate system $\bar Z\equiv(\bar{\bf R}, \bar M, \bar V,\bar\theta)$, where $\bar{\bf R}$ is the gyrocenter position, $\bar M$ is the conserved magnetic moment, $\bar V$ is the parallel speed and $\bar\theta$ is the gyrophase, the equations of motion take the form
\begin{eqnarray}\label{push}
\frac{d\bar{\bf R}}{dt}&=&\bar V{\bf b}+\frac{e_{s}}{m_s\Omega_s}{\bf b}\times\nabla\phi-\frac{\bar V}{m_s\Omega_s}{\bf b}\times\nabla a_{\parallel}\nonumber\\
&&+\left\lbrack\frac{\bar M}{m_s}+\frac{\bar V}{\Omega_{s}}\left(\bar V+\frac{a_{\parallel}}{m_s}\right)\right\rbrack{\bf b}\times\nabla\ln B,\nonumber\\
\frac{d\bar M}{dt}&=&0,\nonumber\\
\frac{d\bar V}{dt}&=&\frac{1}{m_s}{\bf b}\cdot\lbrace\left\lbrack\frac{e_s}{\Omega_{s}}\left(\bar V+\frac{a_{\parallel}}{m_{s}}\right)\nabla\phi+\frac{\bar M}{m_s}\nabla a_{\parallel}\right\rbrack\times\nabla\ln B\nonumber\\
&&+\frac{e_s}{m_s\Omega_s}\nabla a_{\parallel}\times\nabla\phi\rbrace-\frac{\Omega_E\bar M}{m_s}{\bf b}\cdot\nabla\ln B.
\end{eqnarray}
Here, the subscript $s$ denotes either EP or thermal ion species and, using the same notations as in~\cite{briguglio95}, $\Omega_s\equiv e_sB_0/m_sc$ is the corresponding cyclotron frequency. The fluctuating potential $a_{\parallel}$ is related to the poloidal magnetic field stream function $\Psi$ through the relationship $a_{\parallel}=(e_s/c)(R_0/R)\Psi$. The parallel electric field term in the equation for $\bar V$ has been suppressed, neglecting, thus, small resistive corrections to the ideal-MHD parallel Ohm's law. Meanwhile, the pressure tensor can be written, in terms of the gyrocenter coordinates, as
\begin{eqnarray}
\Pi_s(t,{\bf x})=\frac{1}{m_s^2}\int d\bar ZD_{Z_c\rightarrow\bar Z}\bar F_s(t,\bar{\bf R},\bar M,\bar V)\nonumber\\
\left\lbrack\frac{\Omega_s\bar M}{m_s}{\bf I}+{\bf b}{\bf b}\left({\bar V}^2-\frac{\Omega_s\bar M}{m_s}\right)\right\rbrack\delta({\bf x}-{\bf \bar R}),
\end{eqnarray}
where $\bf I$ is the unit tensor, $I_{ij}\equiv\delta_{ij}$, $\bar F_s(t,\bar{\bf R},\bar M,\bar V)$ is the gyrocenter distribution function and $D_{z_c\rightarrow\bar Z}$ is the Jacobian of the transformation from canonical to gyrocenter coordinates. The distribution function $\bar F_s$ satisfies the Vlasov equation
\begin{eqnarray}\label{F}
\left(\frac{\partial}{\partial t}+\frac{d{\bf\bar R}}{dt}\cdot\nabla+\frac{d\bar V}{dt}\frac{\partial}{\partial\bar V}\right)\bar F_s=0,
\end{eqnarray}
where $d\bar{\bf R}/dt$ and $d\bar V/dt$ are given by equation~\ref{push}.
In the numerical implimentation of XHMGC, equations~\ref{push} and \ref{F} can be readily solved as a full-F simulation. On the other hand, a $\delta f$ algorithm~\cite{briguglio98,tajima89,kotschenruether88,parker93} is also implemented in order to minimize the discrete particle noise. The latter is recommended as far as $\delta f\ll\bar{F}$, the former when $\delta f\approx\bar{F}$. 

\section{Applications}\label{app}
In general, the extended version of HMGC can have two species of kinetic particles. On one hand, one can use XHMGC for investigating thermal ion kinetic effects on Alfv\'enic modes driven by EP. On the other hand, it may be interesting to use XHMGC as a tool to simulate two coexisting EP species, generated e.g. by both ion cyclotron resonance heating (ICRH) and neutral beam injection (NBI) heating, in order to study linear excitation of Alfv\'enic fluctuations and Energetic Particle Modes (EPM)~\cite{chen94}, as well as the interplay between the respective nonlinear physics controlled by the different heating sources~\cite{pizzuto10}.

HMGC has been extensively used in~\cite{briguglio95} and~\cite{vlad95} to investigate the linear physics (damping and EP drive mechnisms), and in~\cite{briguglio98} and~\cite{zonca05} to analyze the nonlinear dynamics of EPM. XHMGC has been verified against those previous findings and can recover numerical simulation results in the above studies. Furthermore, by accounting for the kinetic thermal ion effects, XHMGC shows the existence of Kinetic BAE (KBAE) which can be seen as radially trapped eigenstates due to discretization of BAE-SAW continuum by FLR/FOW effects, as well as KBAE resonantly excited by wave-particle interactions with EPs~\cite{wang10a,wang10b}.

\begin{figure}[t]
	\centering
	\includegraphics[width=0.5\textwidth]{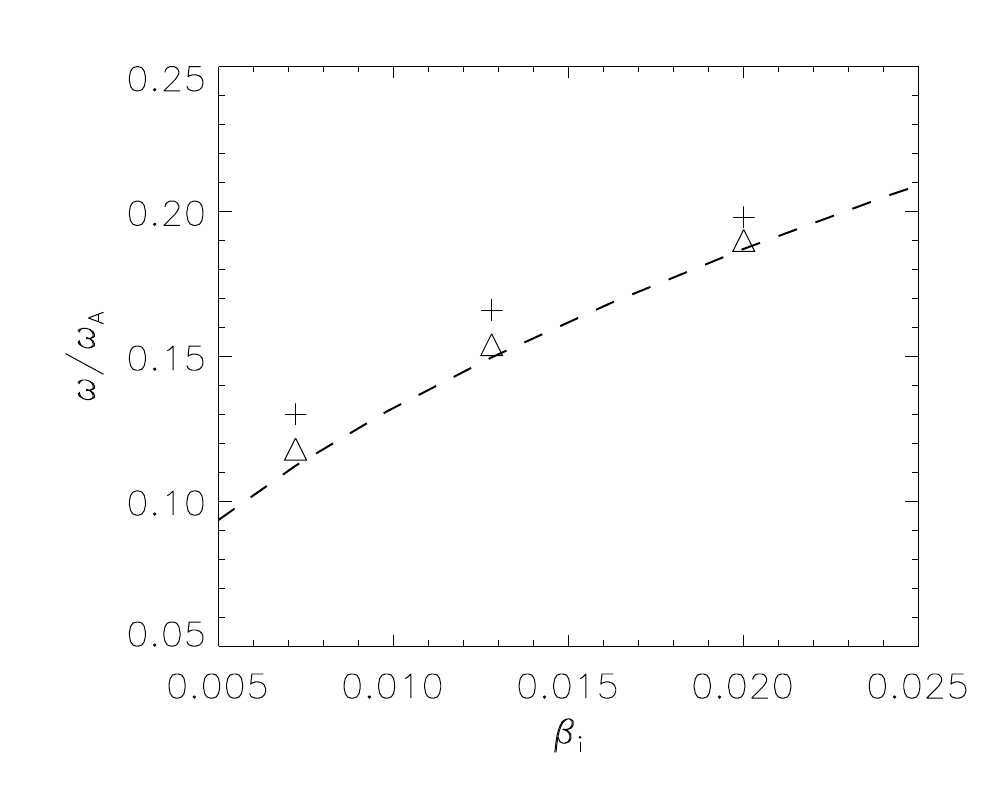}
	\caption{Real frequency comparison between simulations by ``antenna'' excitations and the theoretical accumulation point frequencies. Simulations refer to an equilibrium magnetic field charaterized by shifted circular magnetic surfaces with inverse aspect ratio $a/R_0=0.1$ and the q-profile, in the cylindrical approximation, given by $q(1)=q(0)+[q(1)-q(0)]r^2$, where $r$ is normalized to $a$, $q(0)=2.7$ and $q(1)=3.9$. $\triangle$ is eigenfrequency from simulations for $n=1$, $+$ is eigenfrequency from simulations for $n=3$, the black dashed line is $\omega_{BAE}$ the accumulation point frequency.}
	\label{freq}
\end{figure}

As an example to demonstrate the capability of XHMGC, we briefly report simulation results of KBAE, which is discussed in detail in~\cite{wang10a}. The results show that a fully kinetic treatment of thermal ions is necessary for a proper description of the low frequency Alfv\'enic fluctuation spectrum. By including thermal ion compressibility, our numerical simulations do show the existence of a finite-frequency BAE accumulation point in the SAW continuum, which was demonstrated analytically and numerically using MHD codes~\cite{chu92}. Meanwhile, when effects due to finite ion drift orbit width (FOW) are included, our simulations clearly demonstrate that the BAE-SAW continuum becomes discretized; yielding a series of discrete kinetic eigenmodes with small frequency separation~\cite{wang10a, zonca98}. In figure~\ref{freq}, we have plotted the BAE accumulation frequencies in the fluid limit which are defined as $\omega_{BAE}=q\omega_{ti}(7/4+T_e/T_i)^{1/2}$ with $T_e/T_i\rightarrow0$ in the current case, as well as eigenmode frequencies obtained from simulation results. The analytically predicted KBAE frequencies~\cite{wang10a} are in good agreement with observations from numerical simulations. The results also indicate that FOW kinetic effects increase with the toroidal mode number, as expected~\cite{wang10a,zonca98}.

\begin{figure}[t]
	\centering
	\includegraphics[width=0.4\textwidth]{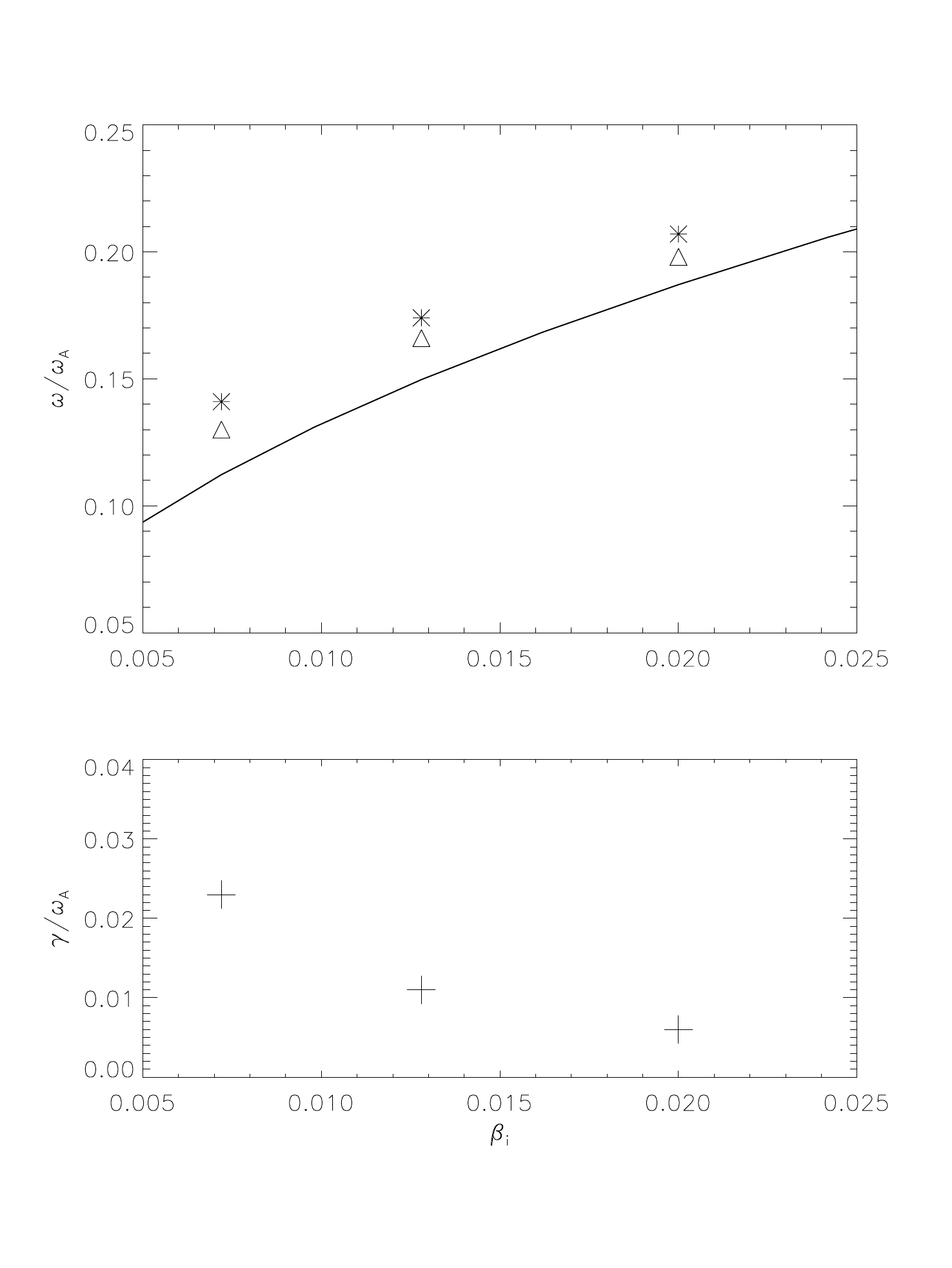}
	\caption{The real frequency $\omega$ and growth rate $\gamma$ for the n=3 mode versus different thermal ion pressure parameters for $\beta_i=0.0072,0.0128,0.02$, and with a fixed value of $\beta_E=0.009$. ``$\ast$'' is the mode real frequency of simulation results by EP excitations; ``$\triangle$'' is the KBAE frequencies by antenna excitations; solid line denotes the theoretical BAE accumulation point frequency; ``$+$'' is the growth rate by EP excitation simulations. }
	\label{epwr}
\end{figure}

On the other hand, our simulations also show that KBAE can be driven by EPs. In figure~\ref{epwr}, we can see that the frequencies scale properly with the KBAE frequencies; and the growth rates decrease with the thermal ion temperature due to the stronger ion Landau damping and/or the weaker EP drive due to the increased frequency mismatch between mode and characteristic EP frequencies. In the absence of thermal ion kinetic effects, the excited modes may be identified as energetic particle mode (EPM); which requires sufficiently strong drive to overcome the SAW continuum damping. Including the thermal ion kinetic effects not only introduce a finite kinetic thermal ion frequency gap at the BAE accumulation frequency but also discretize the BAE-SAW continuum. In that case, the continuum damping is greatly reduced or nullified, and the discrete KBAE's are more readily excited by the EP drive.

\section{Conclusions and discussions}\label{conclusions}
In the present work, we have employed the theoretical framework (generalized kinetic vorticity and quasi-neutrality equations) of the generalized linear fishbone dispersion relation and derived an extended hybrid MHD-gyrokinetic simulation model applicable to the low-frequency regime, where effects of thermal ion compressibility and diamagnetic drifts play significant roles in the dynamics of Alfv\'en waves and energetic particles in tokamak plasmas. The extended simulation model has been implemented into an eXtended version of HMGC (XHMGC). Initial simulations of XHMGC have discovered the existence of KBAE discretized by the thermal ion FOW effects, which are absent in conventional MHD codes. Simulations also demonstrate that KBAE can be readily excited by EPs. In the current model, we have taken $T_e/T_i\rightarrow0$ and neglected finite Larmor radius effects in order to simplify the presentation and focus on  the most important qualitative new physics connected with implementation of the thermal ion compressibility. In addition, XHMGC is limited to circular shifted magnetic surfaces equilibria, with relatively large aspect ratio; XHMGC includes kinetic effects related to both bulk and fast ions; however, it is typically used for retaining only the perturbed pressure for two EP species; XHMGC doesn't include rotation (see \ref{AA}), while it retains the perturbed electrostatic potential. These additional effects will be considered in future works.

More recently, the electromagnetic formulation~\cite{holod09} of global gyrokinetic particle simulation in toroidal geometry has been implemented in GTC~\cite{lin98}. In such a code, ions are treated by the gyrokinetic equation, while electrons are simulated using an improved fluid-kinetic electron model~\cite{holod09}. In~\cite{deng10}, the connection between the extended hybrid MHD-gyrokinetic model and gyrokinetic simulation model has been verified in the drift kinetic limit as well as ignoring the terms on the order of $O((\epsilon/q)^2)$. Instead of directly calculating the pressure tensor, lower moments of the kinetic equation have been calculated, i.e. the perturbed density and parallel current. Using charge neutrality condition, it can be demonstrated that the combination of the perturbed density and parallel current contribution is totally equivalent to the pressure tensor in equation~\ref{eq:twofluidperp2}. Therefore, both GTC and XHMGC can be verified using different models in a common validity regime; yielding more detailed understanding of the underlying physics. 

On the other hand, kinetic compressibility is also included in (linear) spectral codes, such as NOVA-K~\cite{fu93,gorelenkov99} and MARS-K~\cite{liu08}. While in NOVA-K only perturbative treatment of the kinetic effects is considered and continuum damping is not taken into account, as in MARS-K, since both are eigenvalue codes, the spectral approach allows the study of the linear stability of all eigenmodes in a general equilibria; meanwhile, kinetic effects are generally related to bulk plasmas only, although the inclusion of fast ions is quite straightforward. As to other hybrid MHD gyrokinetic codes, M3D~\cite{park92} is based on the pressure coupling equation; MEGA~\cite{todo98,todo05} uses a hybrid model for MHD and energetic particles, where the effect of the energetic ions on the MHD fluid is taken into account in the MHD momentum equation through the energetic ion current. The diamagnetic drift effect is evaluated in the MHD equations by adding the diamagnetic advection term to the equation of motion~\cite{tang80,tang82,sugiyama00,mizuguchi00,khan07}. At present, XHMGC can handle two species kinetic particles self-consistently, but is limited to circular shifted magnetic flux surfaces equilibria with vanishing bulk plasma equilibrium pressure. Meanwhile, a new version of the code with general equilibria is being developed, with the capability of solving fully compressible gyrokinetic particle response.

\section*{Acknowledgments}
This work is supported by the ITER-CN under Grant No.2009GB105005,  the NSF of China under Grant No. 11075140, Euratom Communities under the contract of Association between EURATOM/ENEA, USDOE GRANTS, SciDAC, and GSEP. 
%

\appendix

\section{Simple derivation of model equations}\label{AA}

Adopting a multi-fluid moment description of plasma dynamics, the force balance equation can be written as
\begin{eqnarray}
\rho_b \left(\frac{\partial}{\partial t} + {\bf v}_b \cdot {\bf \nabla} \right) {\bf v}_b + \rho_E \left(\frac{\partial}{\partial t} + {\bf v}_E \cdot {\bf \nabla} \right) {\bf v}_E \nonumber\\
= - {\bf \nabla} P_e - {\bf \nabla} \cdot {\bf P}_i -  {\bf \nabla} \cdot {\bf P}_E + \frac{{\bf J} \times {\bf B}}{c} \;\; . \label{eq:twofluid}
\end{eqnarray}
Here, $\rho_b$ and $\rho_E$ are bulk plasma and EP mass densities, ${\bf v}_b = {\bf b} \times {\bf \nabla} P_{0 i \perp}/(\rho_b \omega_{ci}) + \delta {\bf v}_b$, ${\bf v}_E = {\bf b} \times {\bf \nabla} P_{0 E \perp}/(\rho_E \omega_{cE}) + \delta {\bf v}_b + {\bf b} u_{E \parallel}$ and  $\delta {\bf v}_b = (c/B_0) \delta {\bf E} \times {\bf b}$ from equation (8), having omitted terms that are $O (\omega_{*pE}/\omega_{cE})$ or higher with respect to the RHS. Furthermore, thermal ion and EP pressure tensors on the RHS have to be interpreted as usual, i.e. with the conventional fluid velocity shift in the definition
\begin{equation}
{\bf P}_{s ij} = m_s \int d {\bf v} (v_i - u_{si})(v_j - u_{sj}) f_s \;\; , \label{eq:pressureshifted}
\end{equation}
with $f_s$ the particle distribution function and $u_{si} = \int d {\bf v} v_i f_s /n_s$. When the pressure tensor is computed form the particle distribution function within the gyrokinetic description, some subtleties are connected with the ordering $u_{si}/v_{ts} \approx \rho_{Ls}/L$ in the plane orthogonal to ${\bf b}$, with $\rho_{Ls}$ the Larmor radius of the $s$-species,  $v_{ts}$ its thermal speed and $L$ the characteristic equilibrium radial scale-length. Thus, in the drift-kinetic limit used in this work, ${\bf P}_s = P_{s\perp} {\bf I} + (\hat P_{s\parallel} - P_{s\perp}) {\bf b}{\bf b}$, with ${\bf I}$ the unit diagonal tensor and
\begin{eqnarray}
P_{s\perp}&=& m_s \int d{\bf v} \frac{v_\perp^2}{2} f_s \;\; , \label{eq:pperp} \\
\hat P_{s\parallel}&=& m_s \int d{\bf v} (v_\parallel - u_{s\parallel 0})^2 f_s \;\; . \label{eq:pparallelhat}
\end{eqnarray}
Note the difference between $u_{s\parallel 0} =  \int d {\bf v} v_\parallel F_{0s} /n_{0s}$, used here,  and $u_{s\parallel} = \int d {\bf v} v_\parallel f_s /n_s$, used in equation~\ref{eq:pressureshifted}, $F_{0s}$ being the (slowly evolving) equilibrium particle distribution function. In equation~\ref{eq:twofluid}, we assumed that only EPs can carry significant parallel fluid velocity.

The perpendicular component of equation~\ref{eq:twofluid} can be further simplified, by noting that ${\bf \nabla} {\bf b} = {\bf b} {\bf \kappa} ( 1 + O(\epsilon^2) )$, with $\epsilon=a/R_0$ and $a$ and $R_0$ the tokamak minor and major radii, and that we are using the optimal ordering $|\omega| \approx |\omega_{*pi}| \ll |\omega_{*pE}|$.  
This allows us to rewrite
\begin{equation}
\left[ \rho_E \left(\frac{\partial}{\partial t} + {\bf v}_E \cdot {\bf \nabla} \right) {\bf v}_E \right]_\perp = \frac{{\bf b} \times {\bf \nabla} P_{0 E \perp}}{\omega_{cE}} \cdot {\bf \nabla}{\bf v}_{E \perp} + \rho_E u_{E \parallel}^2 {\bf \kappa} \;\; , \label{eq:approxEP}
\end{equation}
where we have dropped the $\propto (\partial_t + \delta {\bf v}_b \cdot {\bf \nabla}){\bf v}_{E \perp} $ terms, for they are $O(\omega/\omega_{*pE})$ and, similarly, the $\propto u_{E \parallel} {\bf b} \cdot {\bf \nabla} {\bf v}_{E \perp}$ term, since it is $O[(T_i/T_E)^{1/2}]$  -- or, equivalently, $O([n_E/n_b)^{1/2}]$ -- for $\omega \approx \omega_{*pi} \approx \omega_{ti}$; at shorter wavelength or higher frequency, this term would be negligible anyway with respect to the thermal ion inertia response, as negligible would be diamagnetic responses of both EPs and thermal ions. So, equation~\ref{eq:approxEP} well describes the physics we want to incorporate in the present analysis. Recalling the definition of ${\bf P}_s$, we also have 
\begin{equation}
- \left({\bf \nabla} \cdot {\bf P}_s\right)_\perp = - {\bf \nabla}_\perp P_{s\perp} - {\bf \kappa} (\hat P_{s\parallel} - P_{s\perp}) \label{eq:divpperp}
\end{equation}
Thus, the second term on the RHS of equation~\ref{eq:approxEP} can be combined with the $\propto \hat P_{E\parallel}$ term on the RHS of equation~\ref{eq:divpperp}, computed for EPs, and actually be reabsorbed into that (up to the relevant order), provided that the pressure tensor is reinterpreted as  ${\bf P}_s = P_{s\perp} {\bf I} + ( P_{s\parallel} - P_{s\perp}) {\bf b}{\bf b}$, with the ``unshifted'' expression
\begin{equation}
P_{s\parallel} =  m_s \int d{\bf v} v_\parallel^2 f_s \;\; . \label{eq:pparallel}
\end{equation}
replacing the usual definition given in equation~\ref{eq:pparallelhat}. With this convention on the pressure tensor, the perpendicular components of equation~\ref{eq:twofluid} can be rewritten as
\begin{eqnarray}
& & \rho_b \left(\frac{\partial}{\partial t} + {\bf v}_b \cdot {\bf \nabla} \right) {\bf v}_b + \frac{{\bf b} \times {\bf \nabla} P_{0 E \perp}}{\omega_{cE}} \cdot {\bf \nabla}{\bf v}_{E \perp} = \nonumber \\ & & \hspace*{3em} - {\bf \nabla}_\perp P_e - \left( {\bf \nabla} \cdot {\bf P}_i \right)_\perp -  \left( {\bf \nabla} \cdot {\bf P}_E\right)_\perp + \left(\frac{{\bf J} \times {\bf B}}{c} \right)_\perp \;\; . \label{eq:twofluidperpp}
\end{eqnarray}
Actually, equation~\ref{eq:twofluidperpp} can be reduced further when residual terms that are $O (\omega_{*pE}/\omega_{cE})$ or higher with respect to the RHS are omitted, as noted below equation~\ref{eq:twofluid}. In fact, one readily obtains
\begin{eqnarray}
& & \left[ \rho_b \left(\frac{\partial}{\partial t} + {\bf v}_b \cdot {\bf \nabla} \right) + \frac{{\bf b} \times {\bf \nabla} P_{0 E \perp}}{\omega_{cE}} \cdot {\bf \nabla} \right]  \delta {\bf v}_b = \nonumber \\ & & \hspace*{3em} - {\bf \nabla}_\perp P_e - \left( {\bf \nabla} \cdot {\bf P}_i \right)_\perp -  \left( {\bf \nabla} \cdot {\bf P}_E\right)_\perp + \left(\frac{{\bf J} \times {\bf B}}{c} \right)_\perp \;\; . \label{eq:twofluidperp}
\end{eqnarray}
This equation readily reduces to the well-known pressure coupling equation [2], in the limit where thermal ion diamagnetic effects and EP contribution to the divergence of the polarization current are neglected.

\section{Study of term (iv) in equation~\ref{1}}\label{A}
In the low-$\beta$ approximation ($\nabla
\ln B_0\simeq\bf\kappa$),
\begin{equation}\label{A1}
\nabla\times\left(\frac{{\bf b}}{B_0}\right)\cong\frac{2{\bf
b}\times\kappa}{B_0}.
\end{equation}
Meanwhile, in the incompressible limit,
\begin{equation}\label{A2}
\frac{\partial}{\partial t}\delta P_{b}+\delta{\bf v_{b}}\cdot\nabla
P_{0b}=0,
\end{equation}
where
\begin{equation}\label{A3}
\delta{\bf v_{b\perp}}=c\frac{{\bf
B_{0}}\times\nabla_{\perp}\delta\phi}{B_{0}^{2}}.
\end{equation}
Then, with equations~\ref{A1},~\ref{A2} and~\ref{A3}
\begin{eqnarray}
\partial_{t}\left(\nabla\times\frac{{\bf b}}{B_0}\right)\cdot\nabla\delta
P_{b}&=&\frac{2{\bf
b}\times\kappa}{B_{0}}\cdot\nabla\frac{\partial\delta
P_{b}}{\partial t}\nonumber\\
&=&c\frac{2{\bf b}\times\kappa}{B_{0}}\cdot\nabla(-\frac{{\bf
B_{0}}\times\nabla_{\perp}\delta\phi}{B_{0}^{2}}\cdot\nabla
P_{0b})\nonumber\\
&=&c\frac{2{\bf b}\times\kappa}{B_{0}}\cdot\nabla(\frac{{\bf
B_{0}}\times\nabla
P_{0b}}{B_{0}^{2}}\cdot\nabla_{\perp}\delta\phi)\nonumber\\
&=&-\frac{2c}{B_{0}^{2}}{\bf k}\times{\bf b}\cdot\nabla
P_{0b}\Omega_{\kappa}\delta\phi,
\end{eqnarray}
where $\Omega_{\kappa}={\bf k}\times{\bf b}\cdot\kappa$.
\section{Study of term (v) in equation~\ref{1}}\label{B}
Assuming
\begin{equation}
\delta{\bf P_{E}}={{\bf b}{\bf b}}\delta P_{E\parallel}+{({\bf I}-{\bf b}{\bf b})}\delta
P_{E\perp},
\end{equation}
we can show
\begin{eqnarray}
\nabla\cdot{\delta{\bf P}_{E}}
&=&(\delta P_{E\parallel}-\delta P_{E\perp})({{\bf b}\nabla\cdot
{\bf b}+{\bf\kappa}})\nonumber\\
&+& {\bf b}\nabla_{\parallel}(\delta P_{E\parallel}-\delta
P_{E\perp})+{\bf\nabla}\delta P_{E\perp},
\end{eqnarray}
where ${\bf\kappa}={\bf b}\cdot{\bf\nabla}{\bf b}$. Thus,
\begin{equation}\label{B3}
{\bf b}\times\nabla\cdot{{\bf P}_{E}}={\bf b}\times\nabla\delta
P_{E\perp}+(\delta P_{E\parallel}-\delta P_{E\perp}){\bf b}\times{\bf\kappa}.
\end{equation}
Now
\begin{equation}
\nabla\cdot(\frac{{\bf b}}{B_{0}}\times\nabla\delta
P_{E\perp})\cong\frac{2{{\bf b}\times{\bf\kappa}}}{B_{0}}\cdot\nabla\delta
P_{E\perp},
\end{equation}
and
\begin{eqnarray}\label{B55}
\nabla\cdot[(\delta P_{E\parallel}-\delta P_{E\perp})\frac{{\bf b}}{B_{0}}\times{\bf\kappa}]&\cong&\frac{{\bf b\times\kappa}}{B_{0}}\cdot\nabla(\delta
P_{E\parallel}-\delta P_{E\perp}).
\end{eqnarray}
In equation~\ref{B55}, we have used the large aspect ratio assumption, $\epsilon\ll1$, consistent with the reduced MHD description used in HMGC~\cite{briguglio95,briguglio98}. Combining equations~\ref{B3} to~\ref{B55}, we obtain
\begin{eqnarray}
\frac{\partial}{\partial t}\nabla\cdot\left(\frac{{\bf
b}\times(\nabla\cdot{\bf\delta
P_{E}})_{\perp}}{B_{0}}\right)&=&\frac{{\bf
b\times\kappa}}{B_{0}}\cdot\nabla\frac{\partial}{\partial t}(\delta
P_{E\parallel}+\delta P_{E\perp})\nonumber\\
&=&\frac{\omega}{B_{0}}\Omega_{\kappa}(\delta P_{E\parallel}+\delta
P_{E\perp})
\end{eqnarray}
\section{Study of the $\propto\delta K_s$ term in equation~\ref{0616}}\label{C}
Here, we assume the definition of $\delta K_s$~\cite{zonca06}
\begin{equation}
e^{iL_{ks}}\delta K_{s}=\delta
f_{s}-\left(\frac{e}{m}\right)_{s}\left[\frac{\partial
F_{0}}{\partial\varepsilon}\delta\phi-\frac{QF_{0}}{\omega}e^{iL_{k}}J_{0}(k_{\perp}\rho_{s})\delta\psi\right]_{s},
\end{equation}
where $\delta f_s$ is the fluctuating particle distribution function,
, on the RHS, we have dropped all terms $\propto \partial F_{0s}/\partial \mu$, for they generate contributions of higher order in what follows~\cite{zonca06}. Thus, in our treatment, $F_{0s}$ is generally anisotropic, although terms $\propto \partial F_{0s}/\partial \mu$ do not appear explicitly. One then finds
\begin{eqnarray}
\left\langle\frac{4\pi
e_{s}}{k^{2}_{\theta}c^{2}}J_{0}(k_{\perp}\rho_{s})\omega\hat{\omega}_{ds}\delta
K_{s}\right\rangle&=&\underbrace{\left\langle\frac{4\pi
e_{s}}{k^{2}_{\theta}c^{2}}\omega\hat{\omega}_{ds}\delta
f_{s}\right\rangle}_{I}-\underbrace{\left\langle\frac{4\pi
e^{2}_{s}}{k^{2}_{\theta}m_{s}c^{2}}\omega\hat{\omega}_{ds}\frac{\partial
F_{0s}}{\partial\varepsilon}\right\rangle\delta\phi}_{II}\nonumber\\
&&+\underbrace{\left\langle\frac{4\pi
e^{2}_{s}}{k^{2}_{\theta}m_{s}c^{2}}\omega\hat{\omega}_{ds}\frac{QF_{0s}}{\omega}J_{0}^{2}(k_{\perp}\rho_{s})\right\rangle\delta\psi}_{III},
\end{eqnarray}
with $<\cdots>$ denoting velocity integration and $J_0$ is the zero order Bessel function. 

For term (I), we obtain
\begin{eqnarray}\label{C3}
(I)&=&\frac{4\pi\omega}{k_{\theta}^{2}c}\Omega_{\kappa}\langle
m_{s}(\mu+v^{2}_{\parallel}/B)\delta f_{s}\rangle=\frac{4\pi\omega}{k_{\theta}^{2}cB}\Omega_{\kappa}(\delta
P_{s\perp}+\delta P_{s\parallel}),
\end{eqnarray}
where
\begin{equation}
\delta P_{s\perp}=\left\langle\frac{m}{2}v_{\perp}^{2}\delta f_s\right\rangle
\end{equation}
and
\begin{equation}
\delta P_{s\parallel}=\left\langle mv_{\parallel}^{2}\delta f_s\right\rangle
\end{equation}
are, respectively, perturbed perpendicular and parallel pressure.

Meanwhile, for $|k_{\perp}\rho|\ll1$, term (III) can be written as  
\begin{eqnarray}\label{C7}
(III)&=&\left\langle\frac{4\pi
e_{s}}{k_{\theta}^{2}c}(\mu+\frac{v_{\parallel}^{2}}{B})\Omega_\kappa\left(\omega\partial_{\varepsilon}F_{0s}+\frac{m_{s}c}{e_{s}B}({\bf
k\times b})\cdot\nabla
F_{0s}\right)\right\rangle\delta\psi\nonumber\\
&=&-\left \langle \frac{4\pi e_s^2}{k_\theta^2 m_s c^2} \omega \hat\omega_{ds} \frac{\partial F_{0s}}{\partial \varepsilon}\right\rangle \left( \delta \phi - \delta \psi \right)\nonumber\\
&&+\frac{4\pi}{k_{\theta}^{2}B^{2}}\Omega_{\kappa}({\bf
k\times b})\cdot(\nabla P_{0s\perp}+\nabla
P_{0s\parallel})\delta\psi.
\end{eqnarray}

Combining equations~\ref{C3},~\ref{C7} and term (II) with the first term on the RHS of equation~\ref{C7}, we obtain
\begin{eqnarray}
\left\langle\frac{4\pi
e_{s}}{k^{2}_{\theta}c^{2}}\omega\hat{\omega}_{ds}\delta
K_{s}\right\rangle&=&\frac{4\pi\omega}{k^{2}_{\theta}cB}\Omega_{K}(\delta
P_{s\perp}+\delta P_{s\parallel})\nonumber\\
&&-\left \langle \frac{4\pi e_s^2}{k_\theta^2 m_s c^2} \omega \hat\omega_{ds} \frac{\partial F_{0s}}{\partial \varepsilon}\right\rangle \left( \delta \phi - \delta \psi \right)\nonumber\\
&&+\frac{4\pi}{k_{\theta}^{2}B^{2}}\Omega_{K}({\bf
k\times b})\cdot(\nabla P_{0s\perp}+\nabla
P_{0s\parallel})\delta\psi.
\end{eqnarray}

\end{document}